\documentclass[prl,showpacs,twocolumn]{revtex4-1}
\usepackage{amsmath}
\usepackage{amssymb}
\usepackage{graphicx}
\usepackage{subfigure}

\begin{document}
\title{Quantum electrodynamic Aharonov-Bohm effect of charge qubit 
}
\author{Young-Wan Kim}
\author{Kicheon Kang}
\email{kicheon.kang@gmail.com}
\affiliation{Department of physics, Chonnam National University, Gwangju 500-757, Korea}
\date{\today}

\begin{abstract}
We predict that a charge under the influence of a quantum 
electrodynamic potential exhibits a force-free Aharonov-Bohm effect. 
The specific system considered in this study  
is a superconducting charge qubit nonlocally interacting with
a cavity electromagnetic field. We find that this nonlocal interaction
gives rise to remarkable quantum electrodynamic phenomena 
such as vacuum Rabi splitting and oscillation, and Lamb shift, under
the condition that
the qubit is located in an electromagnetic-field-free region. 
Our result can be verified in a realistic experimental setup,
and provides a new approach of investigating the Aharonov-Bohm effect
combined with quantum electrodynamics.
\end{abstract}

\pacs{03.65.Vf, 
      42.50.Pq, 
      74.78.Na} 

\maketitle


%
The Aharonov-Bohm (AB) effect~\cite{ab59,Ehrenberg49,Peshkin89} is a milestone 
in our understanding of the electromagnetic interaction.
The essence of the AB effect is that an observable quantum 
interference occurs even when a charged particle moves in a
electromagnetic-field-free region and does not undergo a local 
interaction with a localized field. 
The AB effect appears as an interference pattern of moving charged particles 
in the presence of the scalar and/or vector potential 
generated by a localized source. The potential in this case is treated as
a classical and static variable.
Interestingly, there is no fundamental reason to limit this force-free quantum 
effect to the case of a
static and classical external potential. 
That is, a charge may interact with a potential produced by a quantized 
electromagnetic field, 
where the field itself does not overlap the charge.
This possibility has rarely been addressed,
and our aim here is to predict that this 
``quantum electrodynamic AB effect" 
can be observed in a realistic system.

Instead of an interferometer with moving charged particles, we consider 
a charge qubit~\cite{Nakamura99,Makhlin01}, which is an artificial
atom in a superposition of two different charge states, interacting with a 
single-mode quantized electromagnetic wave.
This is a solid-state analogue of cavity quantum electrodynamics 
(QED)~\cite{Blais04, Schoelkopf08,You11}.
Standard cavity QED deals with 
the interaction of a real atom with single or multiple mode(s) of quantized
electromagnetic waves confined in a cavity 
(see, e.g., Ref.~\onlinecite{Scully97}).
In this cavity QED, it is typically the atomic electric dipole ($\mathbf{p}$) 
which interacts with the electromagnetic field, and the
interaction Hamiltonian is given by
$-\mathbf{p}\cdot\mathbf{E}$, where $\mathbf{E}$ is the cavity electric field.
Therefore, the atom should be located inside the cavity and make local
interaction with the electromagnetic field, 
in order to provide observable effects.
Notably, this condition is not indispensable for an electric
monopole (charge). In this Letter, we show that, for a charge
qubit located {\em outside} a cavity, an indirect interaction
between the charge and the cavity field occurs via the time-dependent
electromagnetic potential. We derive explicitly 
the interaction Hamiltonian for this system.
This interaction
results in a relative phase shift between the two different charge states
and gives rise to the quantum electrodynamic AB effect.
Further, we show that the interaction can be sufficiently strong 
to be observed even in the vacuum limit of the cavity field. 
Naturally, this interaction effect is pronounced near the resonance of 
the qubit and the cavity mode energies. 
Near the resonance, the charge and the cavity photon are
 nonlocally exchanged by the interaction,
which leads to various interesting quantum electrodynamic phenomena 
such as vacuum Rabi oscillation and shift, 
even in the absence of the direct overlap of the charge and the electromagnetic
field. This dynamic AB effect becomes stronger for a finite number
of photons in the cavity due to the larger effective interaction strength. 

\begin{figure}[htp]
\centering
\includegraphics[height=45mm]{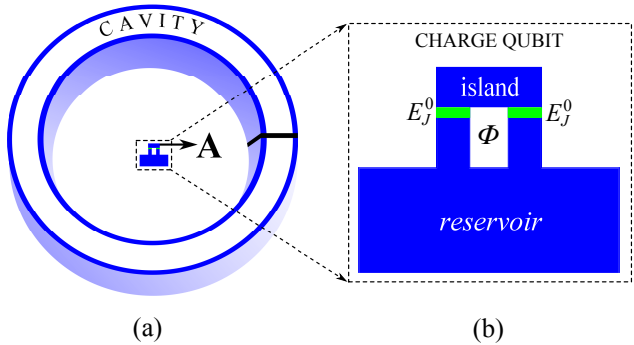}
\caption{(a) Schematic of model system. 
The charge qubit nonlocally interacts with the electromagnetic field 
of a single-mode cavity with a node. 
Note the direction of the vector potential $\mathbf{A}$ generated by
the cavity field.
(b) Magnified qubit diagram.
The qubit is formed by a Cooper pair box tunnel-coupled to a reservoir, 
with its effective coupling
controlled by the external magnetic flux, $\Phi$.
}
\label{fig:Schematic_setup}
\end{figure}

{\em Model.}-
Let us consider the model system illustrated in Fig.~\ref{fig:Schematic_setup}.
A superconducting charge qubit is located in the center of the system
surrounded by a ring-shaped cavity with a node. 
The qubit state is formed via Josephson tunneling with a superconducting 
reservoir~\cite{Nakamura99,Makhlin01}.
The important feature of this setup is that the electromagnetic field
is confined inside the cavity and does not directly overlap the qubit.  

The Lagrangian of the system is given by
\begin{subequations}
\label{eq:lagrangian}
\begin{equation}
 L = L_q + L_c + L_{int} \,,
\end{equation}
where $L_q$, $L_c$, and $L_{int}$ represent the Lagrangians of the 
qubit, the cavity, and
the interaction between the two subsystems, respectively.
In general, the interaction Lagrangian has the Lorentz-covariant form 
\begin{equation}
 L_{int} = \frac{1}{c}\int\mathbf{j}\cdot\mathbf{A}\,d^3\mathbf{r} 
         - \int\rho_q V\, d^3\mathbf{r}
  \,, 
\label{eq:Lint}
\end{equation}
\end{subequations}
where $\mathbf{j}$ and $\rho_q$ are the current and the charge densities 
of the qubit, respectively, interacting with the vector ($\mathbf{A}$) and the
scalar (${V}$) potentials produced by the cavity. 
As we will show below, it is possible to arrange the setup such that 
$\mathbf{j}\cdot\mathbf{A}=0$, so that the magnetic interaction
term vanishes in Eq.~(\ref{eq:Lint}). Therefore, we neglect this 
magnetic interaction hereafter.

The Lagrangian (Eq.~(\ref{eq:lagrangian})) is transformed into the Hamiltonian 
\begin{equation}
H = H_{q} + H_{c} + H_{int} \,,
\end{equation}
where $H_q$, $H_c$, and $H_{int}$ represent the Hamiltonians of the qubit, 
the single mode cavity, 
and their interaction, respectively. The qubit is described by
\begin{subequations}
\begin{equation}
 H_q = \sum_{n=0}^1 \epsilon_n|n\rangle\langle n| 
     - \frac{1}{2} \left( E_J|1\rangle\langle 0| +  E_J^*|0\rangle\langle 1| 
                   \right) \,,
\end{equation}
where $\epsilon_n=E_c(n-n_g)^2$ is the energy of the charge state 
$n$ ($=0,1$), 
with the charging energy $E_c$ of a single Cooper pair. 
This level is tunable via the
gate-dependent parameter $n_g$.
$E_J$ is the effective Josephson coupling, which
can be controlled by an external magnetic flux~\cite{Makhlin01,Makhlin99}.
This qubit Hamiltonian can be diagonalized as
\begin{equation}
 H_q = E_g |g\rangle\langle g| + E_e |e\rangle\langle e| \,,
\end{equation}
where
\begin{eqnarray}
 |g\rangle &=& \cos{(\gamma/2)}|0\rangle + \sin{(\gamma/2)}|1\rangle \,, \\
 |e\rangle &=& -\sin{(\gamma/2)}|0\rangle + \cos{(\gamma/2)}|1\rangle \,,
\end{eqnarray}
\end{subequations}
with the parameter $\gamma$ defined by $\gamma=\tan^{-1}{[E_J/(E_c(1-2n_g))]}$.
The dynamics of the qubit is determined by the difference of the two eigenstate 
energies,
$\hbar\omega_q \equiv E_e-E_g = \sqrt{[E_c(1-2n_g)]^2+E_J^2}$.
The cavity is modeled by a single resonant mode 
%
\begin{equation}
H_c = \hbar\omega \left( a^\dagger a + \frac{1}{2} \right) \,,
\end{equation}
where $a^\dagger$($a$) creates (annihilates) a photon of 
frequency $\omega$,
which is determined by the inner ($\rho_1$) and
the outer ($\rho_2$) radii of the cavity.
An important parameter is the detuning of the two frequencies, which is defined
by $\Delta\equiv\omega_q-\omega$.
Finally, the interaction between the two subsystems is given by
\begin{subequations}
\label{Hint}
\begin{equation}
H_{int} = \hat{q}V .
\label{eq:intHamil}
\end{equation}
The size of the qubit is neglected here. 
The charge operator, 
 $\hat{q} \equiv 2e|1\rangle\langle1|$,
can be written in the eigenstate basis as
\begin{equation}
\hat{q} = e \sin \gamma \left( \sigma_+ + \sigma_- \right),
\label{ChargeOp}
\end{equation}
\end{subequations}
where $\sigma_+=|e\rangle\langle g|$ and $\sigma_-=|g\rangle\langle e|$.
Here, the unimportant energy level shift term is omitted. 

{\em Cavity potentials.}-
The potentials generated by the cavity are obtained as follows.
First, from the cavity geometry~(Fig.~1), one can
find the lowest mode of the cavity electromagnetic field 
\begin{subequations}
\label{eq:QEF}
\begin{eqnarray}
 \mathbf{E}(\rho,\phi,t) &=& \hat{\epsilon} E_0(\rho) 
  \left(a^\dagger e^{i\omega t} + a e^{-i\omega t} \right)   \,, \\
 \mathbf{B}(\rho,\phi,t) &=& i\hat{z} B_0(\rho,\phi) 
  \left( a^\dagger e^{i\omega t} - a e^{-i\omega t} \right) \,,
\label{eq:B}
\end{eqnarray}
where the polarization vector $\hat{\epsilon}$ and the coefficients, 
$E_0(\rho)$ and $B_0(\rho,\phi)$,
are given by 
\begin{eqnarray}
 \hat{\epsilon} &=& - u(\rho) \sin(\phi/2) \hat{\rho} 
       - 2\rho \frac{du}{d\rho} \cos(\phi/2) \hat{\phi} \,, \\
 E_0(\rho) &=& \sqrt{\frac{ \hbar \omega}{2 \beta \rho^2\delta z}} \,, \\
B_0(\rho,\phi) &=& 2E_0(\rho)k\rho u(\rho) \cos \left(\phi/2\right) \,,
\label{eq:B0}
\end{eqnarray}
respectively.
Here, $\beta = k^2\int \rho |u(\rho)|^2\,d\rho$ is a dimensionless factor,
and
the radial function $u(\rho)$ is a combination of the Bessel functions,
 $J_{1/2}$ and $J_{-1/2}$, such that 
\begin{equation}
 u(\rho) = AJ_{1/2}(k\rho) + BJ_{-1/2}(k\rho) .
\end{equation} 
\end{subequations}
The mode wavenumber $k$ and the coefficients $A,B$
are determined by the boundary conditions of the fields. 
Here, $\delta z$ stands for the height of the cavity. 

The electromagnetic field of Eq.~(\ref{eq:QEF}) gives rise to the
oscillating surface charge and
current densities, $\sigma$ and $\mathbf{K}$, respectively, at each
conducting plate 
\begin{subequations}
\begin{eqnarray}
 \sigma &=& \frac{1}{4\pi} \left(\hat{n} \cdot {\bf E} \right),
 \label{SurfaceChargeDensity} \\
 {\bf K} &=& \frac{c}{4\pi} \left( \hat{n} \times {\bf B}  \right),
 \label{SurfaceCurrentDensity}
\end{eqnarray}
\end{subequations}
where $\hat{n}$ is the normal vector to the surface.
The potentials are calculated from these densities 
(with the Lorentz gauge, for convenience)
as integrations over the surfaces ($da'$) of the conductor, with
\begin{subequations}
\begin{eqnarray}
 V({\bf r}, t) &=& \int \frac{\sigma({\bf r'}, t_r)}{|\mathbf{r}-\mathbf{r}'|} 
  \,da'\,, 
 \label{eq:retardedV} \\
 {\bf A}({\bf r}, t) &=& \frac{1}{c} \int 
   \frac{{\bf K}({\bf r'},t_r)}{|\mathbf{r}-\mathbf{r}'|} \,da'\,,
\label{eq:retardedA}
\end{eqnarray}
\end{subequations}
where $t_r \equiv t-|\mathbf{r}-\mathbf{r}'|/c$ is the retarded time, 
implying that the influence of the cavity state propagates with the speed
of light, $c$. 

From Eqs.~\eqref{eq:B},\eqref{eq:B0},\eqref{SurfaceCurrentDensity},
and \eqref{eq:retardedA}, we find that the vector potential at the 
location of the qubit is directed toward
the node (see Fig.\ref{fig:Schematic_setup}). The detailed form of
$\mathbf{A}$ is not important here, as we are interested in the case where
$\mathbf{j}\cdot\mathbf{A}=0$ which can be intentionally established.
In this case, the interaction
between the two subsystems is determined entirely by the scalar potential
$V(t)$ at the location of the qubit. The result is expressed as
\begin{subequations} 
\begin{equation}
 V(t) = V_1(t) + V_2(t) \,,
\end{equation}
where $V_1(t)$ and $V_2(t)$ are the contributions from the surface charges of
the inner and the
outer conductors of the cavity, respectively, given by 
\begin{equation}
 V_j(t) = V_j^0 \left( ae^{-i\omega( t - \rho_j/c)}   
        + a^\dagger e^{i\omega (t - \rho_j/c)} \right) , \;\; (j=1,2), 
\end{equation}
with amplitude
\begin{equation}
 V_j^0 = (-1)^j \frac{\hbar \omega}{|e|} 
  \sqrt{\frac{ \alpha k \delta z}{2\pi^2 \beta }} 
  \frac{u(\rho_j)}{k\rho_j}.
\end{equation}
Here, $\alpha\equiv e^2/\hbar c$ is the fine structure constant. 
Notably, the two contributions, $V_1(t)$ and $V_2(t)$, do not cancel,
despite the fact that the two conducting plates provide 
a vanishing net electromagnetic
field outside the cavity. The potential can be rewritten in the simplified
form
\begin{equation}
 V(t) = - V_0 \left( a e^{-i(\omega t-\theta)} 
      + a^\dagger e^{i(\omega t-\theta)} \right) \,,
\end{equation}
with amplitude $V_0$ and phase shift $\theta$ given by
\begin{equation}
 V_0 = \frac{\hbar \omega}{|e|} \sqrt{ 
       \frac{ D^2 \alpha k\, \delta z}{2\pi^2 \beta }}\,,
\end{equation}
and 
\begin{equation}
 \theta = k\rho_1 
   - \arctan{ 
       \left( 
         \frac{a_2\sin{(k\,\delta\rho)}}{a_1-a_2\cos{(k\,\delta\rho)}} 
       \right) 
             } \,.
\end{equation}
\label{ScalarPot}
Here $\delta\rho\equiv\rho_2-\rho_1$ is the width of the cavity, and
the dimensionless constant
\begin{equation}
 D = \sqrt{a_1^2 + a_2^2 - 2a_1a_2\cos{(k\,\delta\rho)} } ,
\end{equation}
\end{subequations}
is determined by the geometry of the system, with 
$a_1\equiv u(\rho_1)/k\rho_1$ and $a_2\equiv u(\rho_2)/k\rho_2$.
%

%
%

{\em Nonlocal interaction of qubit and cavity field.}-
Combining Eqs.~\eqref{Hint} and \eqref{ScalarPot}, we obtain
\begin{subequations}
\begin{equation}
 H_{int} =  \hbar g(\sigma_++\sigma_-)(a e^{i\theta} + a^\dagger e^{-i\theta}),
\label{eq:Hint}
\end{equation}
where the interaction strength is given by
\begin{equation}
 \hbar g = -\frac{qV_0}{2} \,.
\label{eq:hg}
\end{equation}
\end{subequations}
That is, the non-overlapping qubit and the cavity field interact 
with amplitude $qV_0/2$, and this represents a manifestation of the 
AB effect for a quantum electrodynamic potential.  
Basically, this interaction gives rise to the various interesting phenomena
observed in standard cavity QED, e.g.,
vacuum Rabi splitting and oscillation, and Lamb shift. 
The question here is whether this interaction 
is sufficiently strong for an experimental observation.
In the following, we show that this is indeed the case, and that the nonlocal
interaction of the qubit and the cavity field can yield observable quantum
electrodynamic phenomena.

The coupling strength $g$ is a function of the geometry and is given by
\begin{subequations}
\label{eq:f}
\begin{equation}
 \frac{g}{\omega} = f\left( \frac{\delta\rho}{\rho_1} \right) 
  \sqrt{\frac{\alpha\,\delta z}{\rho_1}} \sin{\gamma} \,,
\end{equation}
where the numerical factor 
\begin{equation}
 f\left( \frac{\delta\rho}{\rho_1} \right) = 
   \sqrt{ \frac{D^2k\rho_1}{2\pi^2\beta} } \,
\end{equation}
\end{subequations}
is plotted in Fig.~2 as a function of $\delta\rho/\rho_1$.
The coupling strength can be estimated in a realistic circuit
QED setup. For example, for a cavity with geometry 
$\rho_1=2.50\mbox{\rm mm}$,
$\delta\rho/\rho_1=0.1$, $\delta z/\rho_1=10^{-3}$, 
and $\sin{\gamma}=1$, we find
$\omega/2\pi=9.09$ {GHz} and $g/2\pi=5.27$ {MHz}. 
This coupling strength is comparable in order of magnitude to
the typical values observed in the
conventional circuit QED setup with a qubit embedded inside a 
cavity~\cite{Schoelkopf08,Wallraff04,Schuster05,Schuster07,Houck07,Fragner08,Paik11}.
This indicates that this nonlocal-interaction-induced effect is measurable. 
We can predict a number of interesting phenomena 
that originate from the
nonlocal exchange of the charge and the cavity photon.
For simplicity, in the following argument, we adopt the rotating wave 
approximation (see e.g., Ref.~\onlinecite{Scully97}), 
where the $\sigma_- a$ and $\sigma_+a^\dagger$
terms are neglected in the interaction Hamiltonian (Eq.~(\ref{eq:Hint})). 
First, the interaction induces vacuum Rabi splitting of the eigenfrequency,
$\omega_\pm = \omega + \Delta/2 \pm \Omega_R/2$, with a vacuum Rabi
frequency of $\Omega_R = \sqrt{4g^2+\Delta^2}$.
The vacuum Rabi splitting and the Lamb shift can be 
observed by controlling
the detuning parameter, $\Delta$, of the qubit and the cavity frequency 
(Fig.~3(a)).
Vacuum Rabi oscillation can also be observed in
a time-domain experiment. At resonance ($\Delta=0$), 
the initially excited qubit exhibits oscillations
in the excited state probability ($P_e$)
of the qubit and in the average photon number ($n_\mathrm{ph}$), as
$P_e(t) = \cos^2( \Omega_R t/2)$ and
$n_\mathrm{ph}(t) = \sin^2(\Omega_R t/2)$ (Fig.~3(b)). 
All of these results are manifestations of the nonlocal interaction of
the charge and the cavity field.

{\em Discussion and outlook.}- 
Several important points concerning our prediction should be further
discussed.
First, the quantum electrodynamic AB effect is not limited to the 
vacuum limit of the cavity. For instance, if the cavity is in a Fock state 
$|n\rangle$ of
$n$ photons, the effective interaction strength becomes stronger. The Rabi 
splitting scales with $\sqrt{n}$, and this causes the effect 
to become more pronounced.

Second, we must check the effect of the local overlap between the charge 
and the 
vacuum electromagnetic field present at the qubit location.
This vacuum field has a continuous spectrum and leads to 
the spontaneous emission of a photon. 
The Hamiltonian of this interaction is given in the form 
\begin{subequations}
\begin{equation}
H_{int}' = \hbar \sum_\mathbf{k} 
 \left(g_\mathbf{k}\sigma_+ a_\mathbf{k} 
     + g_\mathbf{k}^*a_\mathbf{k}^\dagger\sigma_-\right),
\end{equation}
where $a_\mathbf{k}$($a_\mathbf{k}^\dagger$) annihilates(creates) a 
photon with wave vector $\mathbf{k}$. 
The coupling constant is given by 
\begin{equation}
 g_\mathbf{k} = 2i\ell e\omega_q\sin{\gamma} 
  \sqrt{ \frac{\pi}{2V\hbar\omega_\mathbf{k}} } \cos{\theta_\mathbf{k}} \,,
\end{equation}
\end{subequations}
where $V$, $\ell$, and $\theta_\mathbf{k}$ are the volume of the free
space (taken to be infinite), the length of the junction, 
and the angle between the current and the polarization of the continuum mode
of the frequency $\omega_\mathbf{k}$, respectively.
Using the Weisskopf-Wigner theory~\cite{Scully97}, 
we obtain a spontaneous decay rate of 
\begin{equation}
\Gamma = \frac{16\alpha\ell^2\omega_q^3}{3c^2}\sin^2\gamma \,.
\end{equation}
For a set of realistic parameters, i.e., $\omega_q/2\pi=5$ GHz, $\ell=300$ nm,
and $\sin{\gamma}=1$, 
we find that $\Gamma\sim (0.8\; \mathrm{s})^{-1}$. 
This decay time is very long compared to the period of the 
vacuum Rabi oscillation obtained from the interaction expressed in
Eq.~(\ref{eq:hg}).
Therefore, the local overlap of the qubit with the continuous vacuum modes
can be neglected when we deal with the nonlocal overlap with a resonant
cavity. Note that this is not the main source of the spontaneous relaxation
of a realistic qubit composed of an electronic 
circuit~\cite{Lehnert03, Blais04, Wallraff05, Kim08, Kim11, Koch07, Paik11, Rigetti12}. 
In any case, the typical spontaneous relaxation time of the qubit is 
sufficiently long 
for observation of the quantum electrodynamic AB effect.

Third, the AB effect due to the time-dependent potential has been studied 
previously in Ref.~\onlinecite{lee92}, without a strict field-free condition.
In contrast, the novelty of our work is that the electrodynamic AB effect 
is predicted to occur when a charge is located in 
a region free from an electromagnetic field.

Finally, we point out that there is an alternative, more natural, 
way to understand the force-free AB effect, 
which was formulated by one of us~\cite{kang13,kang15}.  
This approach is based on the local interaction of the electromagnetic fields,
and the interaction Lagrangian of Eq.~\eqref{eq:Lint} is replaced by
\begin{equation}
 L_{int} = \frac{1}{4\pi} \int (\mathbf{B}_q\cdot\mathbf{B}
           - \mathbf{E}_q\cdot\mathbf{E}) \,d\mathbf{r}  \,,
\label{eq:Lint2}
\end{equation}
where $\mathbf{E}_q$ and $\mathbf{B}_q$ stand for the electric and the
magnetic fields, respectively, generated by the charge qubit.
The quantum electrodynamic AB effect predicted here can also be derived
from the alternative interaction Lagrangians of Eq.~\eqref{eq:Lint2}.



%

{\em Conclusion.}-
We have predicted the quantum electrodynamic AB effect,
which can be observed in a charge qubit
nonlocally interacting with a cavity field,
under the condition that the charge is placed in a region free from
an electromagnetic field. 
The effect becomes prominent when the qubit level splitting
is in resonance with the frequency of the cavity field. A number of
interesting quantum electrodynamic phenomena are predicted, 
including vacuum Rabi splitting and 
oscillation, and Lamb and Stark shifts, as a result of the nonlocal 
exchange of the charge and the cavity photon.   
Our result shows that the AB effect can be investigated using a very new
approach involving a quantum electrodynamic external potential. 
 
This work was supported by the National Research Foundation of Korea under
Grant No. 2012R1A1A2003957.

%
%
\bibliography{reference-AB}

\begin{figure}[htp]
\includegraphics[height=45mm]{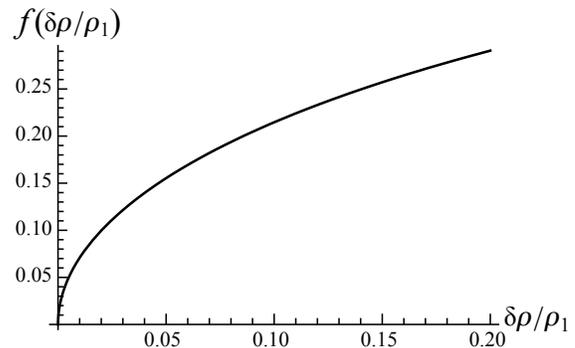}
\caption{Numerical factor $f(\delta\rho/\rho_1)$ of interaction
strength (see Eq.~(\ref{eq:f})) determined by cavity geometry. 
}
\label{fig:NFact}
\end{figure}
\begin{figure}[htp]
\centering
	\subfigure[]{\label{fig:energy}
		\includegraphics[width=70mm]{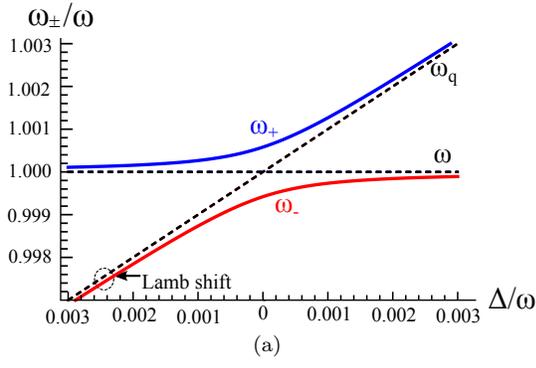}}
	\\
	\subfigure[]{\label{fig:RabiOSC}
		\includegraphics[width=70mm]{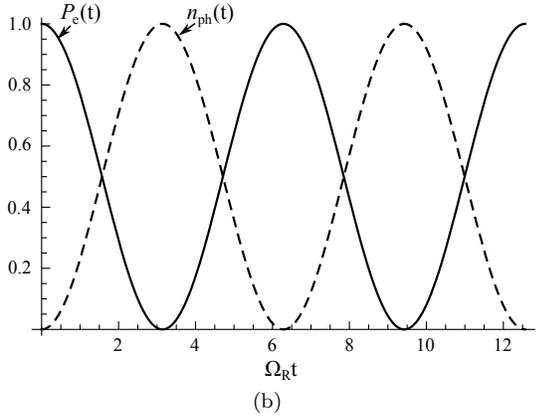}}
		
\caption{Vacuum fluctuation effects driven by nonlocal interaction of 
charge qubit and cavity field for $\delta\rho/\rho_1 =0.1$ 
and $\delta z/\rho_1 = 10^{-3}$.
(a) The vacuum Rabi splitting and the Lamb shift are represented by the dressed
eigenfrequencies $\omega_\pm$ (solid lines) 
as functions of the detuning $\Delta$.
(b) Vacuum Rabi oscillation of the excited state probability of the
qubit ($P_e$) and the average photon number ($n_\mathrm{ph}$) at resonance 
($\Delta=0$).
}
\end{figure}
\end{document}